# Precursor superconducting effects in the optimally doped YBa$_2$Cu$_3$O$_{7-\delta}$ superconductor: the confrontation between superconducting fluctuations and percolative effects revisited

I. F. Llovo[*,1,2], J. Mosqueira[1,2], C. Carballeira[1] and F. Vidal[1]

## Abstract

The confrontation between the superconducting fluctuations and percolation effects as the origin of the in-plane paraconductivity in cuprate superconductors was earlier addressed at a quantitative level in the case of the optimally doped YBa$_2$Cu$_3$O$_{7-\delta}$ (YBCO) compound. Using in-plane resistivity data from a high-quality YBCO thin film, we will extend these analyses to high reduced temperatures, in the case of the Gaussian-Ginzburg–Landau (GGL) approach for the conventional superconducting fluctuations, by considering the total energy cutoff. These data will also be analysed in terms of the mean field-approach of the effective-medium theory, to probe if emergent percolative effects may account for the resistivity rounding above $T_c$. Our analyses confirm earlier conclusions: the measured paraconductivity cannot be explained in terms of emergent percolation processes, but it may be accounted for in terms of the GGL approach. These results also call into question alternative scenarios, including a recent proposal derived from emergent percolative effects.

## Acknowledgements

This work was supported by the Agencia Estatal de Investigación (AEI) and Fondo Europeo de Desarrollo Regional (FEDER) through project PID2019-104296GB-I00, and by Xunta de Galicia (grant GRC no. ED431C 2018/11). I.F. Llovo acknowledges financial support from Xunta de Galicia through grant ED481A-2020/149.

---

[*]Iago Fernández Llovo, iagof.llovo@usc.es | [1]QMatterPhotonics, Departamento de Física de Partículas, Universidade de Santiago de Compostela, E15782 Spain | [2]iMATUS, Universidade de Santiago de Compostela, E15706 Spain

# 1 Introduction

The dilemma between percolation processes and superconducting fluctuations to account for the electrical resistivity rounding observed above but near the critical temperature $T_c$ in cuprate superconductors was already posed by Bednorz and Müller in their seminal work [1]. This dilemma was addressed at a quantitative level in Ref. [2] by studying the dc in-plane resistivity $\rho_{ab}(T)$ of the prototypical optimally doped $YBa_2Cu_3O_{7-\delta}$ (YBCO) superconductor. By using the simplest version of the mean field-approach of the effective-medium theory (EMT) [3], the possible presence of emergent percolative processes was shown to play a negligible role in the measured paraconductivity. In contrast, the rounding could be explained in the so-called mean field region by considering the unavoidable presence of the superconducting fluctuations [2], using the Lawrence-Doniach (LD) approach for layered superconductors of the Gaussian-Ginzburg-Landau (GGL) scenario [4].

Despite the earlier confrontation commented above (see also Refs. [5-7]), the dilemma between superconducting fluctuations and emergent percolation mechanisms to explain the measured paraconductivity in cuprate superconductors is still far from being closed [8-18]. In this paper, previous measurements obtained in high-quality YBCO thin films will be analysed, by comparing the paraconductivity data with conventional superconducting fluctuation models, extended to higher reduced temperatures $\varepsilon = \log(T/T_c)$ with the inclusion of a total energy cutoff in the energy spectrum to account for the high-energy fluctuating modes [19, 20]. Complementarily, in order to probe if emergent percolative effects may account for the resistivity rounding above $T_c$, the data will be analysed in terms of the EMT approach used in Ref. [2], extended now up to the resistivity rounding onset ($T_{onset} \approx 150$ K), determined as the temperature above which ρ(T) follows the normal-state linear behaviour within the noise level.

Our findings evidence that the GGL approach can account for the sudden decrease that paraconductivity experiences at high $\varepsilon$ (for $\varepsilon \gtrsim 0.1$). However, the EMT approach fails to qualitatively explain our data when percolation is considered as the sole cause of paraconductivity. Our results are in stark contrast with recent proposals that postulate emergent percolation as the sole cause of rounding above $T_c$ in underdoped compounds [8].

## 2 Theoretical background

### 2.1 Conventional superconducting fluctuations: Lawrence-Doniach model with a total energy cutoff

Fluctuation effects in superconductors induce the creation of evanescent Cooper pairs above the superconducting transition temperature, which have a measurable effect in different observables. For instance, a progressive decrease of the resistivity is manifested as the temperature approaches $T_c$ from above. In low-$T_c$ superconductors, such a decrease is negligible. In contrast, the small amplitude of the superconducting coherence length in cuprate superconductors (about 1 nm) enhances fluctuation effects, and the change in the resistivity just above $T_c$ is notable (~20% the normal-state resistivity 1 K above $T_c$). The fluctuation-induced electrical conductivity, defined as $\Delta\sigma \equiv \sigma - \sigma_B$ where $\sigma_B$ is the background (normal-state) contribution, can be calculated in the framework of the Lawrence-Doniach model of Josephson coupled superconducting layers. Previous calculations in the framework of the GGL approach for conventional superconducting fluctuations have shown that a so-called total-energy cutoff is required to explain the high reduced temperature paraconductivity data when the high energy modes are considered [19, 20]. For layered materials with strong inter-layer coupling, for which the Lawrence-Doniach (LD) approximation is applicable, the paraconductivity is given by [19],

$$\Delta\sigma^{LD} = \frac{e^2}{16\hbar d_{eff}} \left[ \frac{1}{\varepsilon}\left(1 + \frac{B_{LD}}{\varepsilon}\right)^{-\frac{1}{2}} - \frac{1}{\varepsilon^c}\left(1 + \frac{B_{LD}}{\varepsilon^c}\right)^{-\frac{1}{2}} \right], \qquad (1)$$

where $d_{eff}$ is the *effective* interlayer distance, calculated as the layer periodicity length $d$ divided by the number of layers per periodicity length; $B_{LD} = [2\xi_c(0)/d]^2$ is the Lawrence-Doniach dimensionality parameter, where $\xi_c(0)$ is the coherence length at $T = 0$ K; and $\varepsilon^c$ is the total-energy cutoff parameter, which represents the reduced temperature at which the fluctuations vanish (i.e., the fully normal behaviour is recovered). When $\varepsilon^c \to \infty$, the well-known Lawrence-Doniach result is obtained [4].

### 2.2 Effective-Medium approach for the percolative scenario

As it is well known, the presence of $T_c$ inhomogeneities can significantly alter the properties of cuprate superconductors [21, 22]. The effect of a continuous $T_c$ distribution on the resistivity can be studied through the approach proposed in Ref. [2], which is based on Bruggeman's effective medium theory [3]. This approach assumes that the sample consists of domains (of larger dimensions than the superconducting coherence length amplitude) with $T_c$ values following a Gaussian distribution $G(T_c)$, characterized by its average critical temperature $\overline{T}_c$, and its full width at half-maximum (FWHM) $\Delta T_c$. The electrical conductivity of each domain is assumed to be $\sigma(T, T_c) \to \infty$ if $T < T_c$, and $\sigma = \sigma_B$ when $T > T_c$, where $\sigma_B$ is the background conductivity, obtained by linear extrapolation from the strange metal region of the normal state. The *effective electrical conductivity* $\sigma^e$ can be then obtained by numerically solving

$$\int_0^\infty \frac{\sigma(T, T_c) - \sigma^e(T)}{\sigma(T, T_c) + 2\sigma^e(T)} G(T_c) \, dT_c = 0. \tag{2}$$

## 3 Analysis

As commented in the introduction, the resistivity data obtained on an optimally doped $YBa_2Cu_3O_{7-\delta}$ (YBCO) thin film, previously reported in Ref. [19], will be analysed. These data are shown in Fig. 1. Details of the thin film growth, characterization, and measuring process are summarized in Ref. [19]. Let us just mention that $T_c$ is 91.1 K, and the transition width is as small as 0.6 K. The normal-state in-plane resistivity ratio $\rho(300\ K)/\rho(100\ K)$ is about 3.1, and the extrapolation to $T = 0$ K leads to a small residual value. These results are consistent with previous measurements in high-quality optimally doped YBCO obtained by other groups [23-34].

The resistivity rounding above but near $T_c$ can be clearly observed in Fig. 1. The associated excess electrical conductivity, defined as $\Delta\sigma = \sigma - \sigma_B$, is plotted in Fig. 2 against the reduced temperature $\varepsilon = \log(T/T_c)$ in a convenient log-log scale. The dotted blue line is the best fit of Eq. (2) to the data, in the accessible $\varepsilon$ range (0.01-0.6), with $\overline{T}_c$ and $\Delta T_c$ as free parameters. This leads to $\overline{T}_c = 55.4 \pm 0.8\ K$ and $\Delta T_c = 57.7 \pm 1.1\ K$. This fit does not reproduce the observed $\varepsilon$-dependence of $\Delta\sigma$, particularly in the high $\varepsilon$ range ($\varepsilon \gtrsim 0.1$). Moreover, the resulting $T_c$ distribution is anomalously wide and inconsistent with the experimental data. This can be better observed in Fig. 3, where the temperatures at which the percolation fraction would be attained (between $p_c \sim 0.15$ and 0.30 [8, 35]) are substantially below the temperature at which $\rho = 0$ is observed. A similar analysis was performed with 2D EMT [3], to explore if the anisotropic nature of these materials could have an influence in the distribution $G(\overline{T}_c, \Delta T_c)$ and in the percolation thresholds. However, the differences are absorbed by the free parameters, obtaining equally bad fits to our data. The corresponding $T_c$ distribution obtained from the best fit to 2D EMT is also anomalously wide: $\overline{T}_c = 63.5 \pm 0.6$ K and $\Delta T_c = 54.2 \pm 1.0$ K.

Additionally, a comparison with the Lawrence-Doniach model, summarized in 2.1, is also included in Fig. 2 (solid red line). The description of this analysis was already presented in Ref. [19]. Let us just mention that the agreement with the experimental data is excellent in the accessible $\varepsilon$ range, leaving the transverse coherence length amplitude $\xi_c(0)$ as the only free parameter, which results to be about 0.1 nm, as expected for optimally doped YBCO. Other parameters in Eq. (2), as $T_c$ or the cutoff constant $\varepsilon^c$, were obtained from the resistive transition midpoint and from the temperature onset of the resistivity rounding, respectively. Finally, it is worth noting that the conventional LD model without a cutoff (dot-dashed black line), fails to describe the data at high reduced temperatures, where the contribution of high-energy fluctuation modes is relevant.

# 4 Conclusions

In this work, the earlier analyses of the possible contributions of emergent percolative effects to the rounding above $T_c$ in optimally doped YBCO [2] have been extended to the high reduced temperature region. Our analysis was performed by comparing data from a high-quality thin film sample to both a percolative model using effective-medium theory (EMT), and the Gaussian-Ginzburg-Landau (GGL) approach for layered superconductors, with the inclusion of a total energy cutoff [19, 20]. We found that extreme $T_c$ distributions are required to try to explain the data in the EMT scenario, as it was previously reported for underdoped compounds [8]. However, the optimally doped sample used in this paper has excellent stoichiometric homogeneity. Additionally, the percolation fraction predicted by EMT for this compound is substantially lower that the expected value $p_c \approx 0.15 - 0.3$ [8, 35]. More worryingly, the existence of regions with $T_c$ as high as 120 K and as low as absolute zero, which the EMT model requires, have not been observed in previous magnetic transition measurements. Therefore, the excess conductivity observed for optimally doped YBCO cannot be explained by effects of emergent percolation alone. Nevertheless, this result does not exclude that $T_c$ distributions may play a role in the paraconductivity of heavily underdoped compounds, as stoichiometric homogeneity cannot be ensured for these compounds, and wider transitions are generally observed. On the other hand, the GGL phenomenological approach offers a much better fit of the data, when a total-energy cutoff is introduced to account for the high-energy fluctuating modes. This approach successfully accounts for the sudden decrease of paraconductivity at high $\varepsilon$. Our results provide quantitative confirmation of the earlier conclusions of Ref. [2], which are now extended to high reduced temperatures: conventional fluctuations alone can quantitatively account for the optimally doped YBCO paraconductivity data both close and far from $T_c$, and that the emergence of percolation must be a second order effect in the resistive rounding above $T_c$ of samples of such stoichiometric quality.

# Appendix: numerical methods

The effective medium theory model was calculated with the self-consistent approach used in Ref. 2. In order to solve Eq. (2) for each temperature $T$, the integral was split in two terms,

$$\int_0^T G(\overline{T}_c, \Delta T_c)\, dT_c + \int_T^{\overline{T}_c + 4\Delta T_c} \frac{\sigma(T, T_c) - \sigma^e(T)}{\sigma(T, T_c) + 2\sigma^e(T)} G(\overline{T}_c, \Delta T_c)\, dT_c = 0\,, \quad (3)$$

to avoid numerical divergences when $\sigma(T, T_c) \to \infty$. All the calculations were performed using Python *scipy* library. The integrals were solved by numerical quadrature integration. Eq. (3) was numerically solved to obtain the values of $\sigma^e$ for any given $\overline{T}_c$ and $\Delta T_c$ pair. The resulting curves were then minimized against the free parameters $\overline{T}_c$ and $\Delta T_c$, to obtain the gaussian distribution $G(\overline{T}_c, \Delta T_c)$ with mean value $\overline{T}_c$ and full width at half maximum $\Delta T_c$ which best explains the data. The results of Eq. (3) were fit to the paraconductivity data obtained with the procedure described in Section 3.

# Conflict of interest

On behalf of all authors, the corresponding author states that there is no conflict of interest.

# Statements and declarations

## Competing Interests

The authors have no relevant financial or non-financial interests to disclose.

## Author contributions

F. Vidal and J. Mosqueira conceptualised the study. I.F. Llovo and J. Mosqueira performed the data analyses.

C. Carballeira helped writing the manuscript. All authors read and approved the final manuscript.

## Data availability

Data will be made available on reasonable request

# Figures

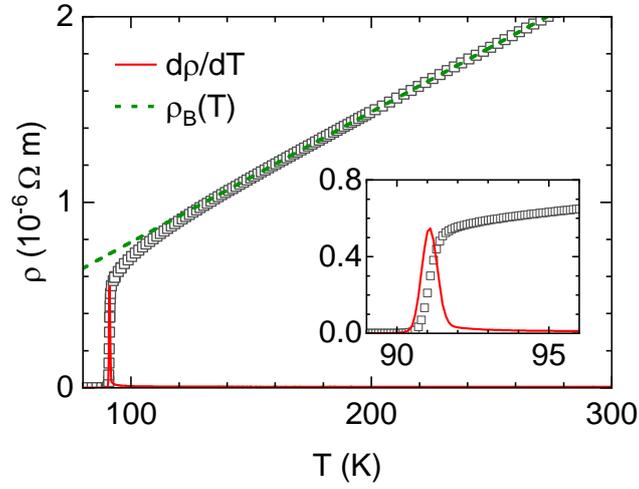

**Fig. 1** In-plane resistivity data $\rho_{ab}(T)$ (black squares) used for the analyses in this work, corresponding to measurements of Ref. [19] in a high quality, optimally doped YBCO thin film. The normal state background resistivity $\rho_B(T)$ (dashed green line) was estimated by using a linear fit of the data between 200 K and 250 K ($2.2 - 2.7\ T_c$), well into the strange metal regime of the normal state (i.e., above $T_{\text{onset}} \approx 1.7\ T_c$). A sharp superconducting transition occurs at $T = 91.1$ K, with FWHM $\Delta T_c = 0.6$ K, as evidenced by the derivative of data (solid red line), also shown in the inset. See Ref. [19] for more details

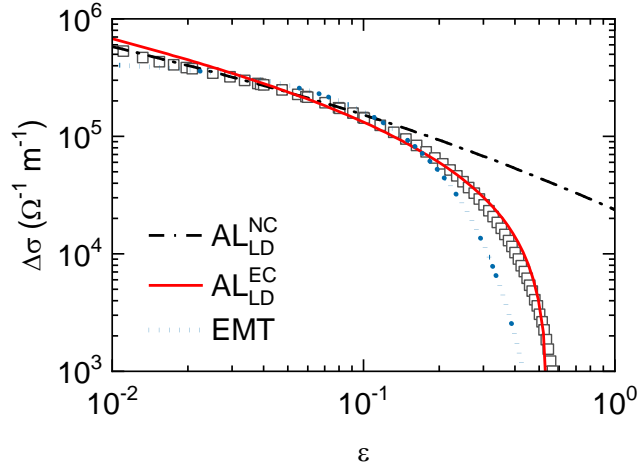

**Fig. 2** Excess conductivity data (black squares) compared with LD and EMT approaches: In the case of the LD approach, $B_{LD}$ is the only free parameter. The solid red line was obtained by imposing $\varepsilon^c = 0.55$ in Eq. (2), as predicted in Ref. [20], and leads to $B_{LD} = 0.15$, whereas the dot-dashed black line was obtained without a cutoff (imposing $\varepsilon^c \rightarrow \infty$ in Eq. (2)) and leads to $B_{LD} = 0.19$. The best fit to the EMT approach, with both $\overline{T}_c$ and $\Delta T_c$ as free parameters, is shown as a dotted blue line. While the LD approach with total energy cutoff for the superconducting fluctuation scenario successfully explains the measurements, the EMT approach for the percolative scenario cannot even qualitatively account for our data, mainly in the high reduced temperature region

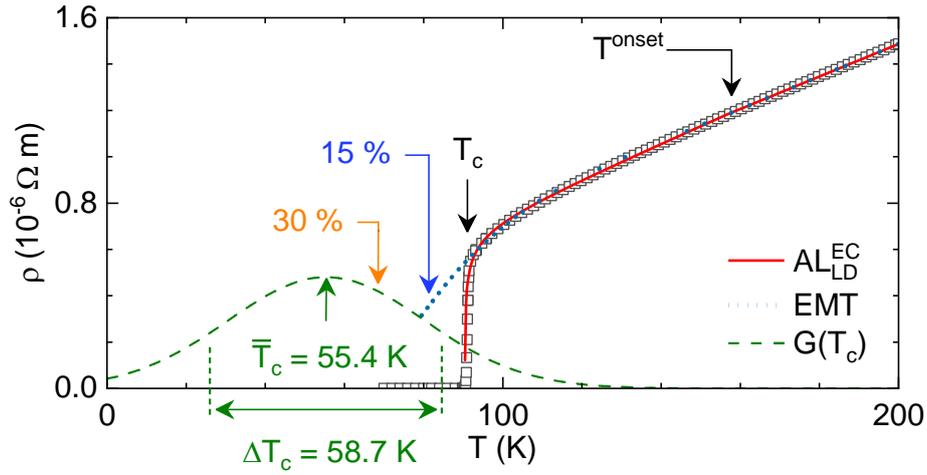

**Fig. 3** Best fits of LD with total energy cutoff (solid red line) and EMT (dotted blue line) approaches to the in-plane resistivity $\rho_{ab}(T)$ data. The EMT scenario fails to faithfully represent the transition both close and far from $T_c$ (see also Fig. 2). The curve representing the distribution G($T_c$) resulting from the best fit is shown for comparison ($\bar{T}_c = 55.4$ K, $\Delta T_c = 57.7$ K, dashed green line). As it can be seen, the $T_c$ distributions that are required to try to explain the data are extremely wide, as previously reported for underdoped compounds [8]. However, the percolation fractions $p_c = 0.15$ and $p_c = 0.3$ [8, 35] that have been proposed for such percolative effects do not appear anywhere near $T_c$ for optimally doped samples. Moreover, the existence of regions with $T_c$ as high as 120 K and all the way down to absolute zero, predicted for YBCO by this model, is not corroborated by previous magnetic transition measurements [19]